\documentclass[aps,prl,twocolumn,groupedaddress]{revtex4}

\usepackage{graphicx}
\usepackage{bm}
\usepackage{graphicx}
\usepackage{subfigure}
\usepackage{latexsym}
\usepackage{placeins}
\usepackage{amstext,amssymb,amsfonts}
\usepackage{epsfig}
\usepackage{color}
\usepackage{titlesec}
\usepackage{amsmath}

\begin{document}

\title{Optimal cost for strengthening or destroying a given network}
\author{Amikam Patron$^1$, Reuven Cohen$^2$, Daqing Li$^{3,4}$ and Shlomo Havlin$^1$}

\affiliation{1 Department of Physics, Bar-Ilan University, Ramat-Gan 5290002, Israel\\
  2 Department of Mathematics, Bar-Ilan University, Ramat-Gan 5290002, Israel\\
  3 School of Reliability and Systems Engineering, Beihang University - Beijing 100191, China\\
	4 Science and Technology on Reliability and Environmental Engineering Laboratory - Beijing 100191, China
 }

\date{\today}

\begin{abstract}
Strengthening or destroying a network is a very important issue in designing resilient networks or in planning attacks against networks including planning strategies to immunize a network against diseases, viruses etc.. Here we develop a method for strengthening or destroying a random network with a minimum cost. We assume a correlation between the cost required to strengthen or destroy a node and the degree of the node. Accordingly, we define a cost function $c(k)$, which is the cost of strengthening or destroying a node with degree $k$. 
Using the degrees $k$ in a network and the cost function $c(k)$, we develop a method for defining a list of priorities of degrees, and for choosing the right group of degrees to be strengthened or destroyed that minimizes the total price of strengthening or destroying the entire network.
We find that the list of priorities of degrees is universal and independent of the network's degree distribution, for all kinds of random networks. 
The list of priorities is the same for both strengthening a network and for destroying a network with minimum cost. However, in spite of this similarity there is a difference between their $p_c$ - the critical fraction of nodes that has to be functional, to guarantee the existence of a giant component in the network.

\end{abstract}

\maketitle

\section{Introduction}
Random networks are obtained by randomly linking a set of nodes by edges. There are many types of random networks, each of them is generated by a specific method and has a typical topology. One of the most important characteristics of networks is the probability distribution of the number of edges that emanate from a randomly chosen node, called the \emph{degree} of the node, and denoted by $p(k)$.

Two kinds of random networks are widely studied, are the Erd\H{o}s-R\'{e}nyi network (ER) and Scale-Free network (SF). In ER network, that was the first model of random networks \cite{erdds1959random,erd6s1960evolution}, as the total number of nodes $N$ tends to infinity, the degrees of the nodes $k$ are distributed according to a Poisson distribution $p(k)=e^{-\lambda}\frac{\lambda^k}{k!}$, where $\lambda$ is the expectation of the node's degree. In a SF network, as $N$ tends to infinity the degrees of the nodes are distributed according to power-law distribution $p(k)=Ck^{-\gamma}$, where $C$ is a normalization factor. In a SF network, although most of the degrees  are relatively small, there is a significant probability for the existence of nodes with high degree, called \emph{'hubs'}, as opposed to ER networks.  

There are two typical states in random networks. The first state is when the network is fragmented into many small components, each of them contains relatively small number of nodes. The second state is when a \emph{'giant component'} exists in the network, which is a component that contains a finite fraction of the entire network's nodes i.e. scales as $O(N)$. The transition between the two states when the giant component appears in the network, is called \emph{percolation transition} of the network.

It was shown \cite{molloy-randstruct-1995} that in a random network, generated by the configuration model, a percolation transition occurs \cite{albert-rev-mdn-phys-2002,newman-networks-2010} at – 
\begin{equation}	
\label{eq:threshold kappa}
\kappa=\frac{\langle k^2\rangle}{\langle k\rangle}=2\enspace.
\end{equation}
where $\langle k\rangle$ and $\langle k^2\rangle$ are the expectations of the degree and the square of the degree of a node in the network, respectively. For $\kappa>2$ the network is in the supercritical region where a giant component exists, and otherwise if $\kappa<2$ the network is in the subcritical region and a giant component does not exist.

An important case, that is treated in this paper, is when a network in the supercritical state is under attack, where nodes (or edges) are destroyed. As long as the giant component still exists, the network is considered to be functional. However, when a critical fraction of nodes (or edges) are destroyed, a phase transition occurs, the giant component collapses into many small components and the network is considered to be nonfunctional. If we define $p$ to be the probability that a randomly chosen node in the network is not attacked, then the fraction of nodes that are not attacked at the threshold between the supercritical and the subcritical states, is called the critical threshold $p_c$ - the probability that a randomly chosen node in the critical state, is functional.  

Research has been focused on three types of attacks against networks - \emph{random attack},  \emph{targeted attack} and \emph{localized attack}. In a random attack the attacker has no information about the network, its topology or characteristics. A fraction of nodes are chosen randomly to be destroyed \cite{albert-nature-2000,cohen-prl-2000,callaway-prl-2000}. On the other hand, in a targeted attack the attacker has some information about the topology and the nodes of the network, and by this determines which nodes to attack and in which order. In a localized attack, just a certain region in the network is affected. The attack begins against one node, and then it spreads over its neighbors and its neighbors of the neighbors etc., until a certain fraction, $1-p$, of the network is removed \cite{shao-njp-2015}. 
Using Eq. (\ref{eq:threshold kappa}) an expression for $p_c$ for random attack was derived \cite{cohen-prl-2000}
\begin{equation}
p_c=\frac{1}{\kappa_0-1}\enspace,
\end{equation}
where $\kappa_0$ is the value of $\kappa$ before the attack begun. From this follows that for ER network under random attack $p_c=1/\lambda$. 

Furthermore, by this criterion it was shown that for SF random networks under random attack, for $\gamma>3$, $p_c$ equals to finite non-zero value, but for $\gamma\leq3$, $p_c$ approaches zero as $N$ approaches to infinity \cite{cohen-prl-2000}. This means that although almost all the nodes in the network are randomly removed, the network still possesses a giant component and is regarded functional.

Under a targeted attack, it was shown that ER networks behave similarly to their behavior under random attack. That is because most of the degrees in the network are close to $\lambda$. Thus choosing nodes randomly or targeting nodes with high degree are not significantly different. In contrast, it was shown that $p_c$ in SF network under targeted attack can be high \cite{callaway-prl-2000,cohen-prl-2001}. This means that removing small fraction of nodes, causes the network to collapse and the giant component to disappear. 
This is explained by the existence of a small fraction of hubs in SF networks that are critical for the connectivity of the network. When the hubs are destroyed, the network breaks into small components.

Although there exists extensive study about attacks against random networks, most of the studies are based on some ideal assumptions. First, it is usually assumed that there are no constraints to be considered by the attacker, like limited budget to the execution of the attack, limited time to the execution of the attack etc.. Furthermore it is assumed that the attack is implemented ideally, such that in a targeted attack the attacker has some information about the network, and in a random attack the attacker knows nothing about the network. Indeed, there are few studies that consider variations on the ideal models of attacks against networks \cite{gallos-prl-2005,schneider-procacascience-2011,achard-journal-neuroscince-2006,li-pre-2007}.

Recently Morone and Makse \cite{morone-nature-2015} developed a method for optimal percolation in random networks. Their method identify the minimal set of nodes that would break the network into disconnected small components without a giant component.   
Also in \cite{Braunstein-pnas-2016} the problem of network dismantling was studied, where the case of random sparse graphs was mapped to the network decycling problem, and an efficient algorithm was presented for finding the minimal set of nodes to be removed and dismantle the network.
But again, these studies do not take into consideration cost constraints on which the attacker is subjected.

In this paper we present an optimized approach for strengthening or attacking a network, where we consider the constraint of minimizing of the cost of strengthening or destroying (which is equal to immunizing) the network. We develop an analytical strategy for choosing the right set of degrees that would strengthen or immunizing the network with minimum cost. Surprisingly, as long as the network is random, the method and the set of degrees are general and do not depend on the degree distribution.

\section{Efficient destruction of a network}
\subsection{Theory}
We begin with a functional network in the supercritical region. We assume a realistic feature of dependency between the cost of destroying or immunizing a node and its degree. Accordingly, we define a cost function $c(k)$ that represents the cost of destroying or immunizing one node with degree $k$. Our goal is to find for every group of nodes with degree $k$, the fraction of nodes to be destroyed (or immunized), that will be denoted by $r(k)$, such that the total cost to destroy (immunize) the network is minimal.

We define a function $P$, that is the total cost to fragment the network, as follows
\begin{equation}
\label{eq:price}
P=\sum_{k=o}^\infty p(k)Nc(k)r(k)\enspace.
\end{equation}
Every attack begins when all the nodes are functional. That means that initially $r(k)=0$ for all the degrees $k$, and obviously $P=0$. During the attack, when nodes with degree $k$ fail, $r(k)$ increases, as does the cost $P$. After the destruction of a sufficient number of nodes, the condition for critical percolation is achieved, the giant component is fragmented and the attack ends. Our goal is to minimize the total cost $P$ for the entire attack.

Eq. (\ref{eq:threshold kappa}), which is the condition for percolation transition, can be written as
\begin{equation}
\label{eq:threshold expectation}
\sum_{k=0}^\infty (k-1)\frac{kp(k)q(k)}{\lambda}=1\enspace,
\end{equation}
where $q(k)$ is the probability that a randomly chosen node with degree $k$ is functional, and $\lambda$ is the original mean degree in the network. Eq. (\ref{eq:threshold expectation}) defines the percolation threshold of a random network when the expectation of the number of edges that emanate from a populated node, reached by following a randomly chosen edge, equals $1$.
Since $q(k)=1-r(k)$ we get 
\begin{equation}
\label{eq:perc_cond_dest}
\sum_{k=0}^\infty (k-1)\frac{kp(k)\left[1-r(k)\right]}{\lambda}=1\enspace.
\end{equation}
Eq. (\ref{eq:perc_cond_dest}) is equivalent to
\begin{equation}
\label{eq:perc_thresh}
\sum_{k=0}^\infty (k-1)\frac{kp(k)r(k)}{\lambda}=\frac{\langle k^2\rangle}{\langle k\rangle}-2\enspace,
\end{equation}
where for each $k$, $0\leq r(k)\leq1$.

According to Eq. (\ref{eq:perc_thresh}) we define a parameter $a(k)$, that represents the contribution to the progress of achieving the condition for the percolation threshold, when attacking (immunizing) all the nodes with degree $k$ $\left(r\left(k\right)=1\right)$, to be:
\begin{equation}
\label{eq:a_k}
a(k)\equiv (k-1)\frac{kp(k)}{\lambda}\enspace.
\end{equation}

According to Eq. (\ref{eq:price}), we define a parameter $e(k)$ that represents the total cost of destroying all the nodes with degree $k$,
\begin{equation}
e(k)\equiv p(k)c(k)N\enspace.
\end{equation}

Next, we define an efficiency parameter $z(k)$ to be the ratio between $a(k)$ and $e(k)$ after neglecting the constants $\lambda$ and $N$, that is the ratio between the contribution of all the nodes with degree $k$ to the destruction of the network and the price for destroying all the nodes with degree $k$,
\begin{equation}
\label{eq:z_k}
\frac{a(k)}{e(k)}\propto\frac{(k-1)k}{c(k)}\equiv z(k)\enspace.
\end{equation}

Since by definition the relation of the contribution to the destruction of the network per cost of removing all the nodes with degree $k$ is maximal when $z(k)$ is maximal, therefore clearly we prefer to destroy degrees with largest efficiency, that is with highest values of $z(k)$.
Accordingly, we define the following method for destroying (immunizing) a network with minimum cost:

(i) For each degree $k$ calculate $z(k)$. 

(ii) Choose degrees to be attacked (immunized) according to the value of $z(k)$ in descending order.
 
The process destroying the network should be stopped when a sufficient amount of chosen degrees are collected, such that if all these nodes are removed, the condition for percolation threshold, Eq. (\ref{eq:perc_thresh}), would be achieved.
The degrees that were chosen would be fully removed (immunized) $(r(k)=1)$, except the last chosen degree that could be attacked partially. Using Eq. (\ref{eq:perc_thresh}) and (\ref{eq:a_k}), the fraction of nodes that would be removed from the last chosen degree is
\begin{equation}
r(k)=\frac{\kappa-2-\sum_i {a(i)}}{a(k)}\enspace,
\end{equation}
where the summation is over all the degrees that were fully removed.

The main point of our method is the behavior of the function $z(k)$, that gives the priority of each degree $k$ to be destroyed relative to the other degrees. 
Analysis of the behavior of $z(k)$ by identifying first the most preferable degree from which the choice of degrees starts, is the $k$ of the extremum maximum point of $z(k)$. This can be implemented by zeroing the first derivative of $z(k)$ (Eq. (\ref{eq:z_k})) and conditioning the second derivative of $z(k)$ to be negative. Since we are interested in $z(k)$ only for the range $2\leq k<\infty$ (destroying a node with degree $0$ or $1$ contributes nothing to the destruction of the network), we find that for a given cost function $c(k)$ with its specific parameters and constants, not in all cases there exists an extremum maximum point within the bounds $2\leq k<\infty$. In other cases there exist only a superior value of $z(k)$ at the bounds or even outside the bounds of that range, either at $k=\infty$ or at $k\leq2$. In these last cases $z(k)$ at $k\geq2$ is a monotonic increasing function or a monotonic decreasing function, respectively.

Accordingly, we demonstrate our analysis of the behavior of $z(k)$ for two functional forms of $c(k)$: (i) the cost function $c(k)$ is a power-law $k^\alpha$ and (ii) the cost function $c(k)$ is an exponential $e^{\beta k}$. Since it is reasonable to assume that as the degree of a node increases the cost required to destroy it rises, we demonstrate here the behavior of $z(k)$ only when $\alpha>0$ and $\beta>0$.

For case (i) it is easy to see that when $0<\alpha\leq2$, $z(k)$ is a monotonic increasing function, and when $\alpha\geq3$, $z(k)$ is a monotonic decreasing function. In the intermediate range $2<\alpha<3$, $z(k)$ is an extremum maximum function (see Appendix B for a detailed computations). Therefore, when $0<\alpha\leq2$ the optimal strategy of attack (removing or immunizing) should be from the high degrees to low, when $\alpha\geq3$ we begin with removing the low degrees, and when $2<\alpha<3$ we remove the intermediate degrees according to descending order of $z(k)$. 
 
(ii) When the cost function, $c(k)$, is exponential $e^{\beta k}$, and when $\beta>0$, we find that when $\beta\geq1.5$, $z(k)$ is a monotonic decreasing function, and when $0<\beta<1.5$, $z(k)$ is an extremum maximum function (see Appendix C for a detailed computations). Therefore when $\beta\geq1.5$ we begin with removing the low degrees, and when $0<\beta<1.5$ we remove the intermediate degrees according to descending order of $z(k)$. Although analytically there is no $\beta$ for which $z(k)$ is a monotonic increasing function, we analyzed also the case of $\beta$ approaching to $0^+$, where the maximum point at $z(k)$ tends to $k$ approaching infinity. In this case, and in fact in every case when $\beta$ is relative small such that the maximum point at $z(k)$ is greater than the maximum degree of the network, $z(k)$ can be considered as a monotonic increasing function. 

Examples for the behavior of $z(k)$ for power-law and exponential cost functions in the various regions discussed above, are illustrated in Fig. 1.

\begin{figure}[htbp]
\begin{center}
 \hspace{-1.34cm}\includegraphics[width=10cm]{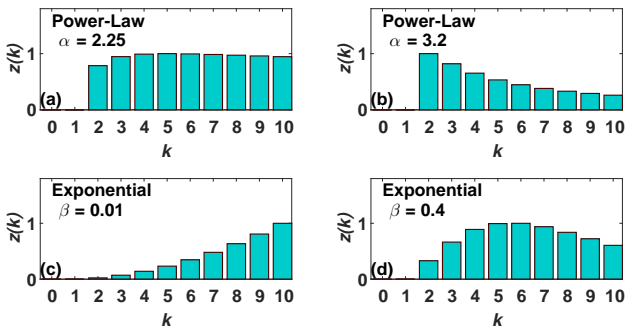}
\end{center}
\caption{\textbf{Demonstrating the behavior of $z(k)$ when the maximum degree of the network is $10$, (a,b) with power-law cost function and (c,d) with exponential cost function: (a)} An extremum maximum for $z(k)$ at intermediate values of $k$ ($k_{max}=5$) when $\alpha=2.25$ (and in general for any $2<\alpha<3$). \textbf{(b)} Monotonic decreasing function of $z(k)$ for $\alpha=3.2$ (and in general for any $\alpha\geq3$). \textbf{(c)} The case of very small $\beta$ where $z(k)$ behaves as a monotonic increasing function. \textbf{(d)} An extremum maximum for $z(k)$ at intermediate values of $k$ ($k_{max}=6$) when $\beta=0.4$ (and in general for all $0<\beta<1.5$). In all cases, for simplifying the demonstration, the values of $z(k)$ were normalized by dividing them by the maximum value of the original series $z(k)$. }
\label{fig:Z_vs_k}
\end{figure}
As shown in Fig. \ref{fig:Z_vs_k} with accordance to Eq. (\ref{eq:z_k}), $z(k)=0$ when $k$ is $0$ or $1$,
that means that nodes with degrees $0$ or $1$ are with the lowest preference to be destroyed (and in fact they should not be destroyed), which is in accordance to the fact that destroying nodes with degrees $0$ or $1$ contributes nothing to the destruction of the giant component and the entire network.

\subsection{Results}
\begin{figure*}[htbp]
\begin{center}
\begin{tabular}{cc}
  \hspace{-7em} \includegraphics[width=10cm]{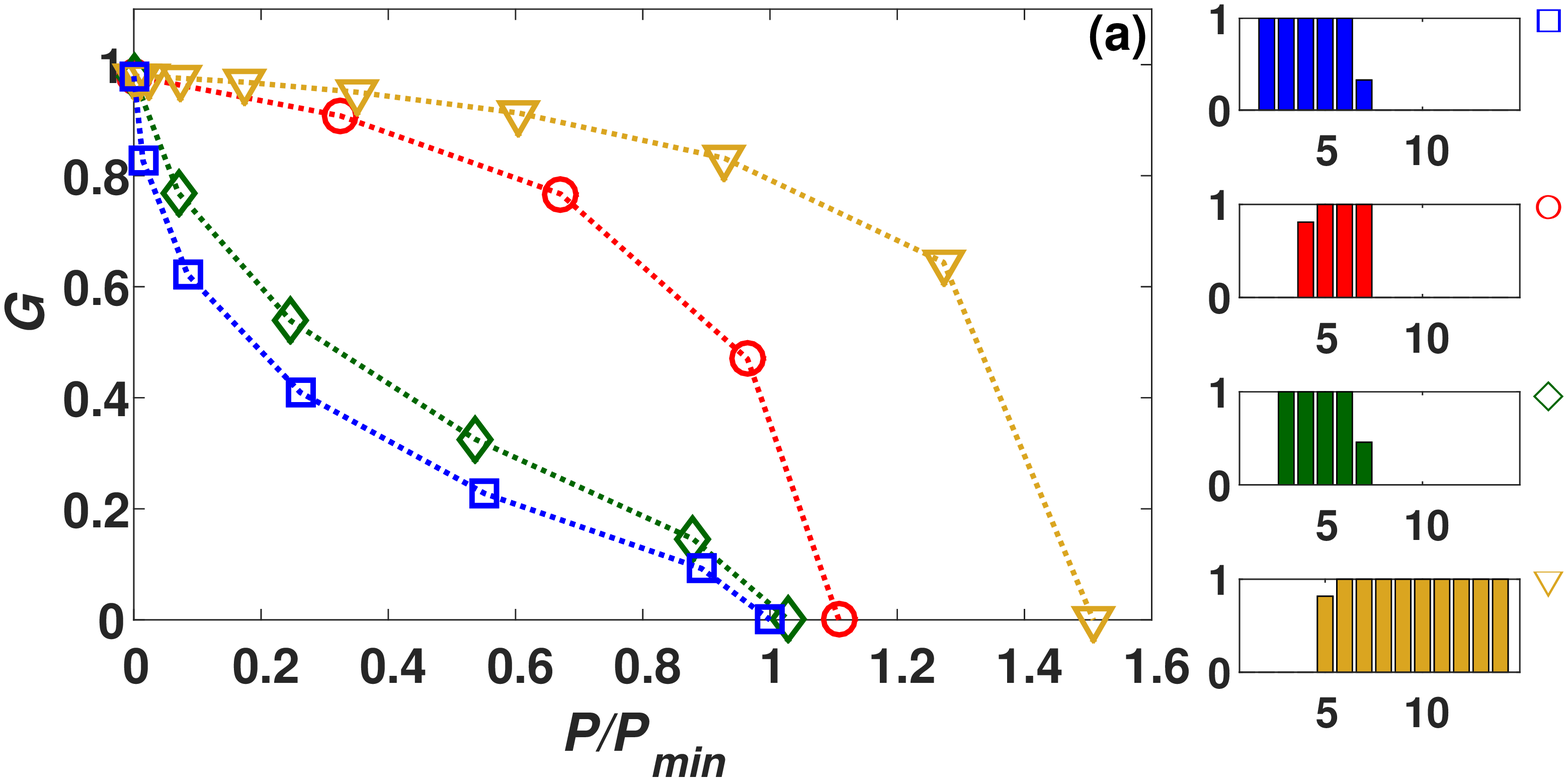}\hspace{-2.5em} &
	\includegraphics[width=10cm]{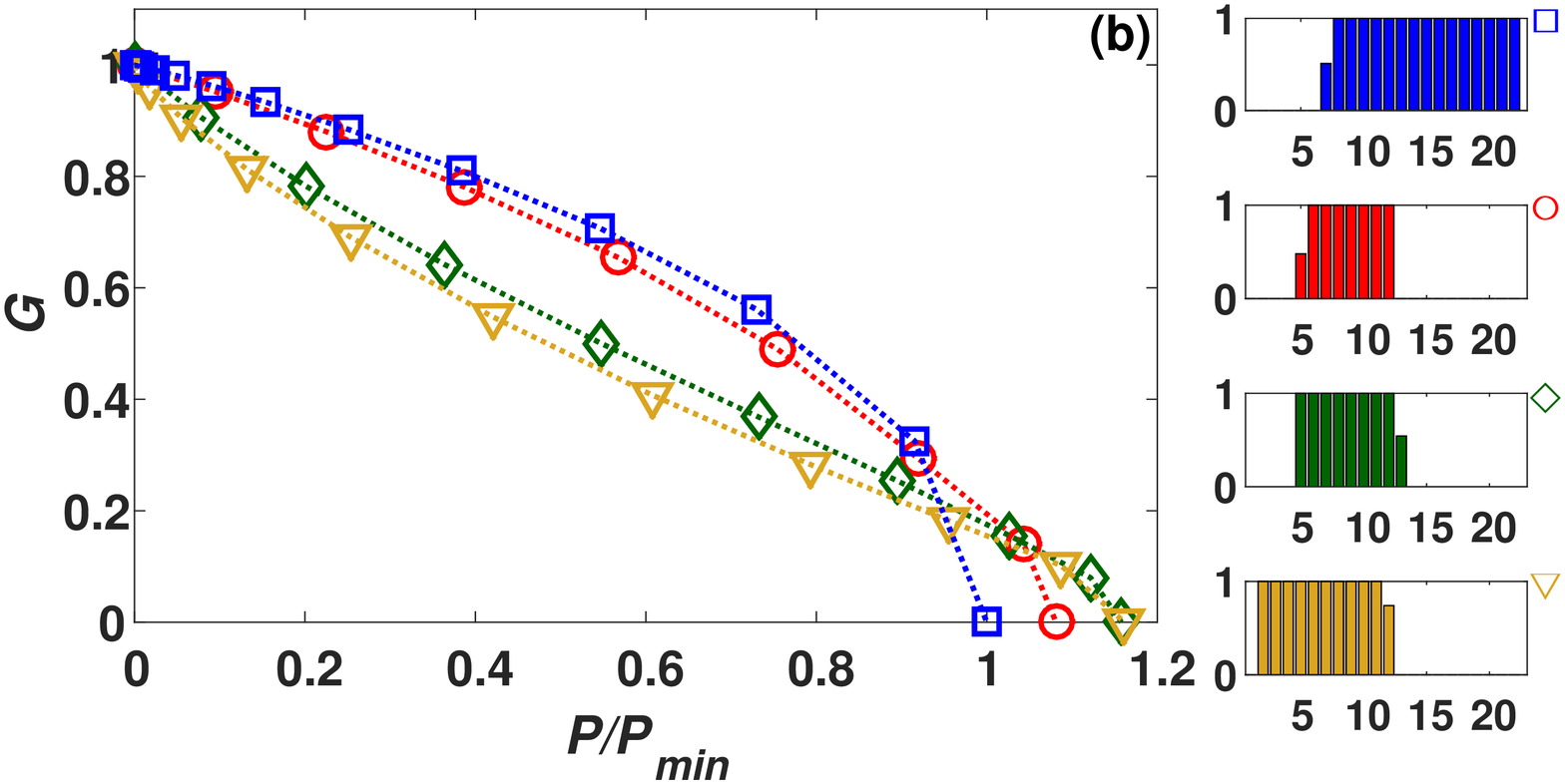} \\
	\hspace{-7em} \includegraphics[width=10cm]{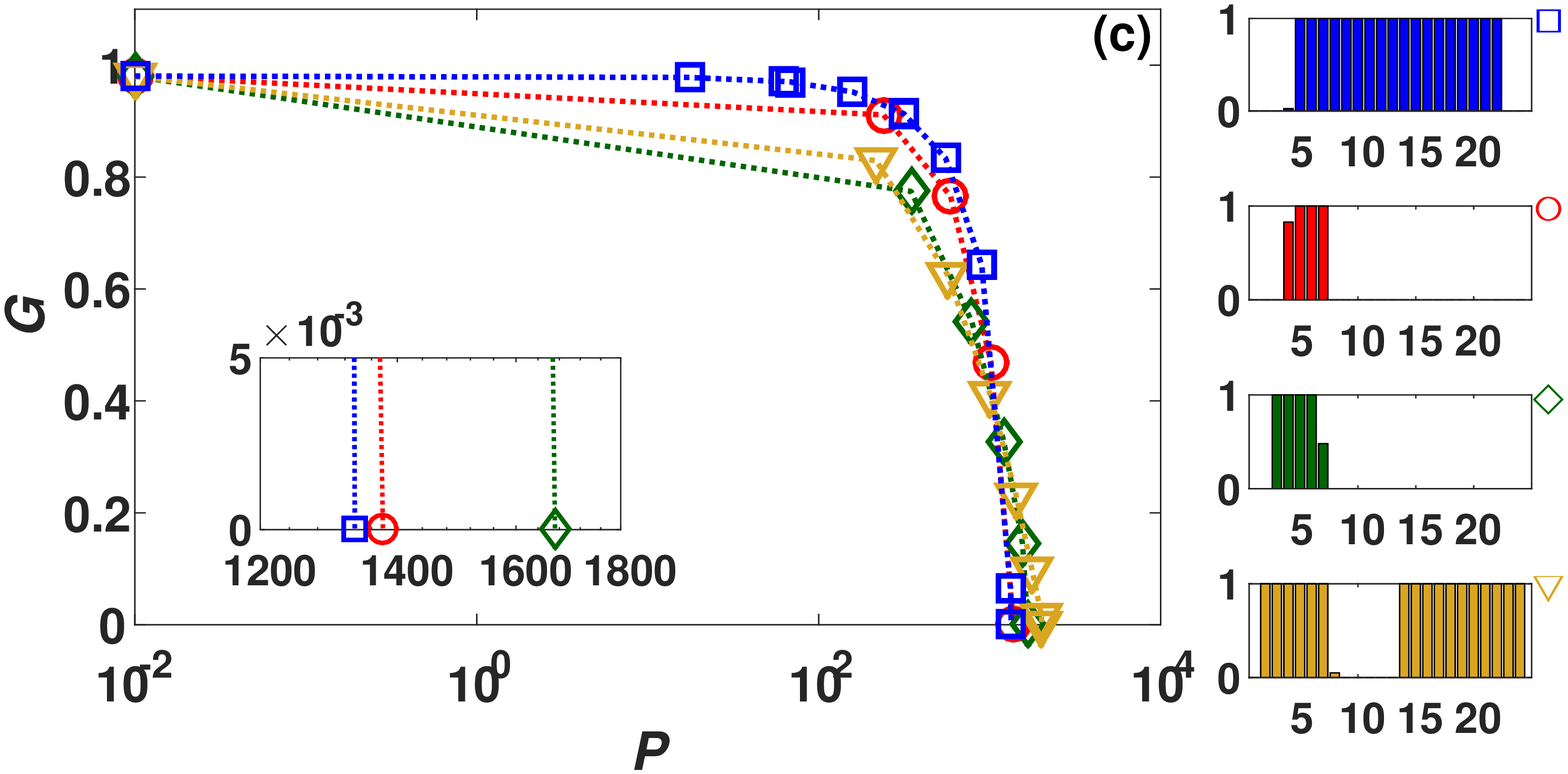}\hspace{-2.5em} &
	\includegraphics[width=10cm]{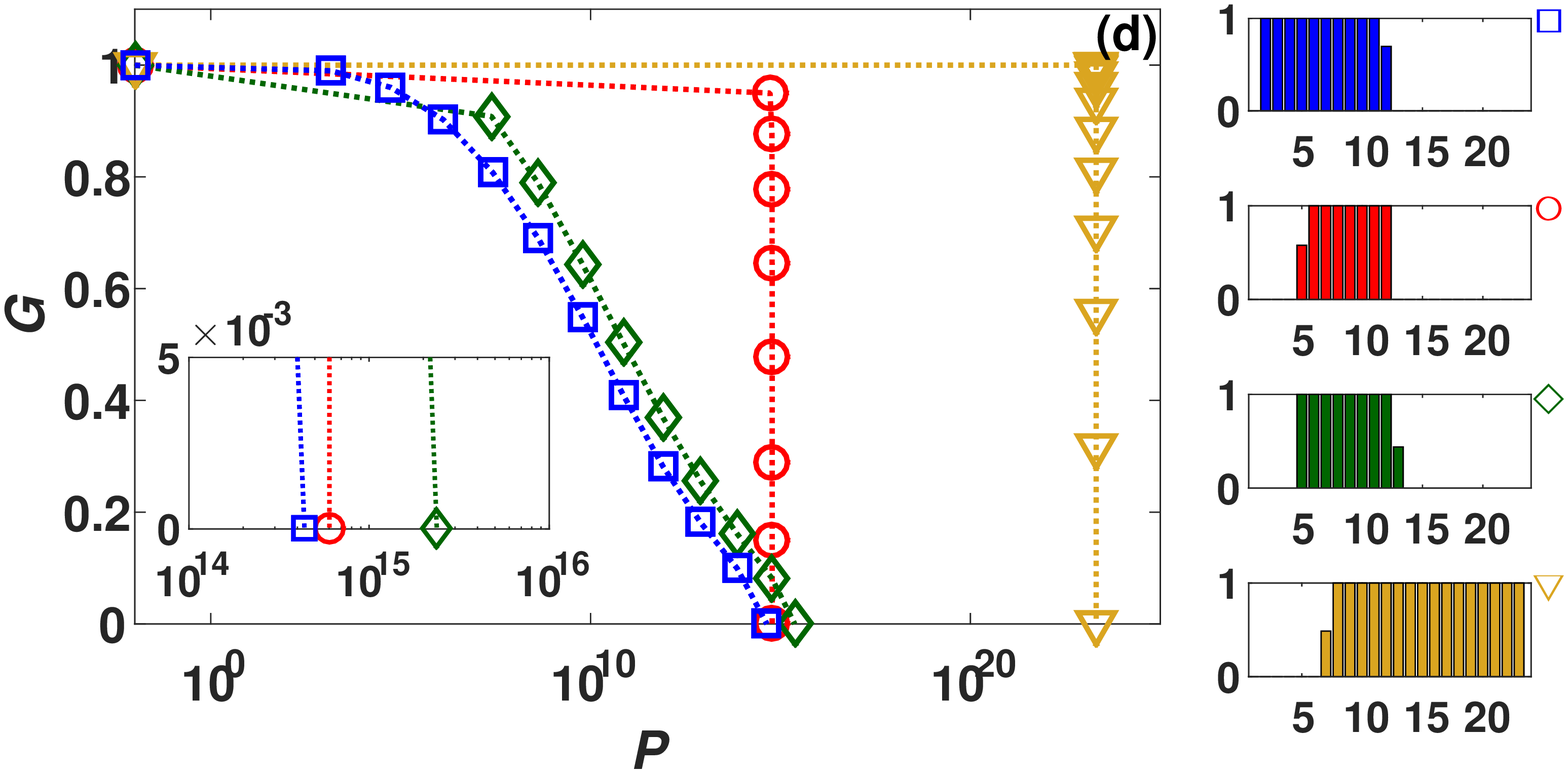} \\
\end{tabular}
\end{center}
\caption{\textbf{(a,b) Behavior of the size of the giant component vs. the normalized accumulated cost:} Cost function is power-law $c(k)=k^\alpha$. Open symbols represent simulations results of ER networks having $N=10^3$ nodes: \textbf{(a)} Average degree $\lambda=4$ and $\alpha=3.2$. \textbf{(b)} Average degree $\lambda=8$ and $\alpha=1$. A normalization of the cost values $P$ on the x-axis was implemented by dividing by the minimum total cost $P_{min}$ of the four curves (strategies).\textbf{(c,d) Behavior of the size of the giant component vs. the cost:} Cost function is exponential $c(k)=e^{\beta k}$: \textbf{(c)} Average degree $\lambda=4$ and $\beta=0.2$. \textbf{(d)} Average degree $\lambda=8$ and $\beta=2.5$. Averages are taken over 50 realizations. The $\alpha$'s and $\beta$'s values were taken as an examples to ranges that state different orders of degrees to be destroyed - descending order from high degrees, ascending order from low degrees and the intermediate degrees.
In the right hand-side of the graphs - the four bars demonstrate the fraction of the removed nodes from each degree, in each graph. Each bar is signed by a symbol respective to the symbol of the graph that it represents. 
Inset in \textbf{(c)} and \textbf{(d)}: Close up of the region where there are several curves that intersect the x-axis very closely. With accordance to our prediction, the curves that represent our method (rectangles) intersect the x-axis at the lowest point compared to the other curves.} 
\label{fig:ER_G_vs_P}
\end{figure*}

The first demonstration of the validity of the theory, in the case of an attack (or immunization) of ER network, is illustrated in Fig. \ref{fig:ER_G_vs_P}. In Fig. \ref{fig:ER_G_vs_P}(a)-\ref{fig:ER_G_vs_P}(b), the cost function is power-law. The graphs present the size of the giant component $G$ vs. the normalized accumulated cost $P/P_{min}$, where $P_{min}$ is the minimal cost of the four strategies discussed below. In Fig. \ref{fig:ER_G_vs_P}(c)-\ref{fig:ER_G_vs_P}(d), the cost function is exponential. The graphs present the size of the giant component $G$ vs. the accumulated cost $P$ on a logarithmic scale (because of the large values of $P$).
In each of the figures \ref{fig:ER_G_vs_P}(a)-\ref{fig:ER_G_vs_P}(d), there are four graphs that illustrate four different attack strategies against the network by a removal of a specific group of degrees. Each choice of the specific degrees for each group, is implemented step by step according to a specific strategy of priorities, and is ended when a sufficient combination of degrees are collected such that if all these nodes are removed the condition for percolation threshold would be achieved. The strategies are: Rectangles - in accordance with our optimal method i.e. according to descending order of the values of $z(k)$ in Eq. (\ref{eq:z_k}). Circles - in descending order from high degrees, except removing the most highest degrees that include $5$ percents of the nodes of the network (because of the high cost of removing very high degrees). Diamonds - in ascending order from low degrees, except removing the most lowest degrees that include $5$ percents of the nodes of the network (because of the negligible contribution to the destruction of the network when removing very low degrees). Downward triangles - an inverse order to the order presented by the theory. 

The beginning point of each graph is its intersection with the y-axis, that represents the initial state before the attack where the size of the giant component is maximal and the accumulated cost is 0. The adjacent point to the beginning point represents the first stage of the attack, where a removal of all the nodes with the most preferred degree, is implemented. That causes the size of the giant component to decrease and the accumulated cost to increase. The next adjacent point in the graph represents the second stage of the attack where all the nodes with the next preferred degree are removed, and the size of the giant component once again decreases and the cost once again increases, and so on. The end of the attack is when a sufficient amount of nodes are removed that causes the giant component to be fully fragmented, is represented by the intersection of the graph with the x-axis where the size of the giant component is 0. The value of the x-coordinate in this point represents the total cost required in order to destroy the entire network. 
We can see in each of the four figures \ref{fig:ER_G_vs_P}(a)-\ref{fig:ER_G_vs_P}(d), that among the four curves, the minimum cost of destroying the network is in the choice that is signed by rectangles that represents the choice of degrees according to our theory, Eq. (\ref{eq:z_k}). Note the interesting case in Fig. \ref{fig:ER_G_vs_P}(c) when it is preferable to destroy the intermediate degrees, as predicted by the theory when $0<\beta<1.5$.

\begin{figure*}[htbp]
\begin{center}
\begin{tabular}{cc}
  \hspace{-7em} \includegraphics[width=10cm]{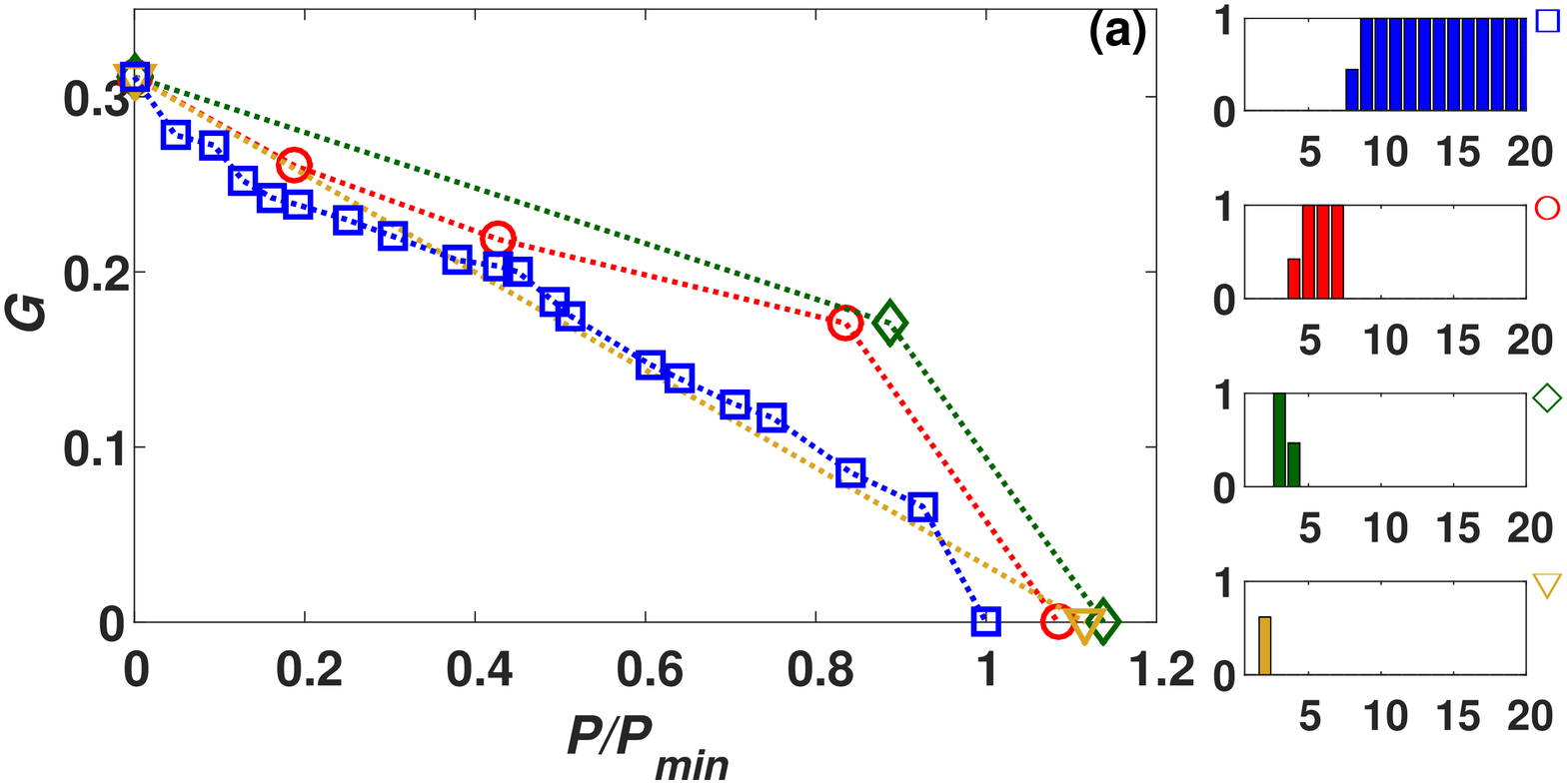}\hspace{-2.5em}  
 & \includegraphics[width=10cm]{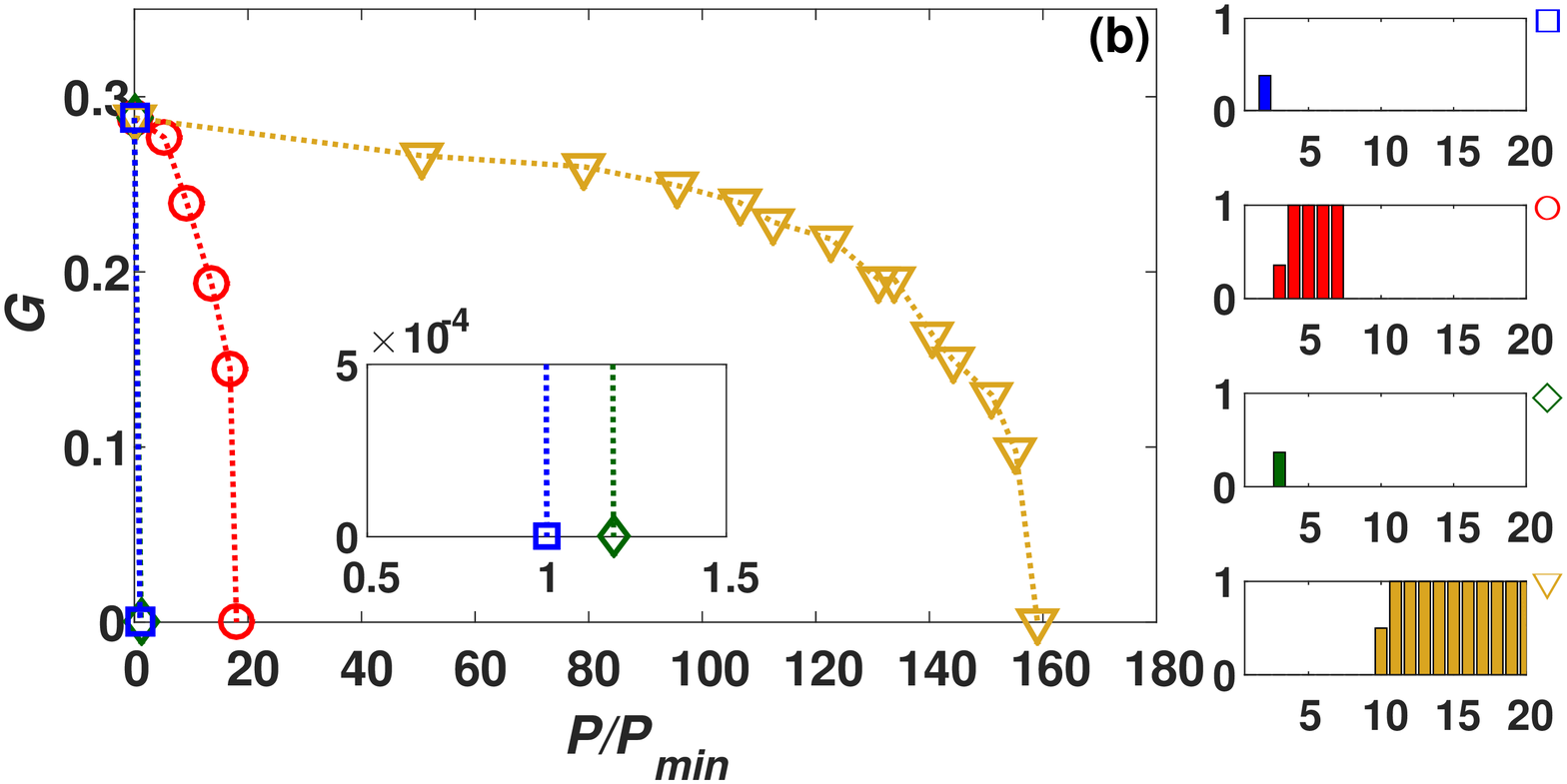} \\
	\hspace{-7em} \includegraphics[width=10cm]{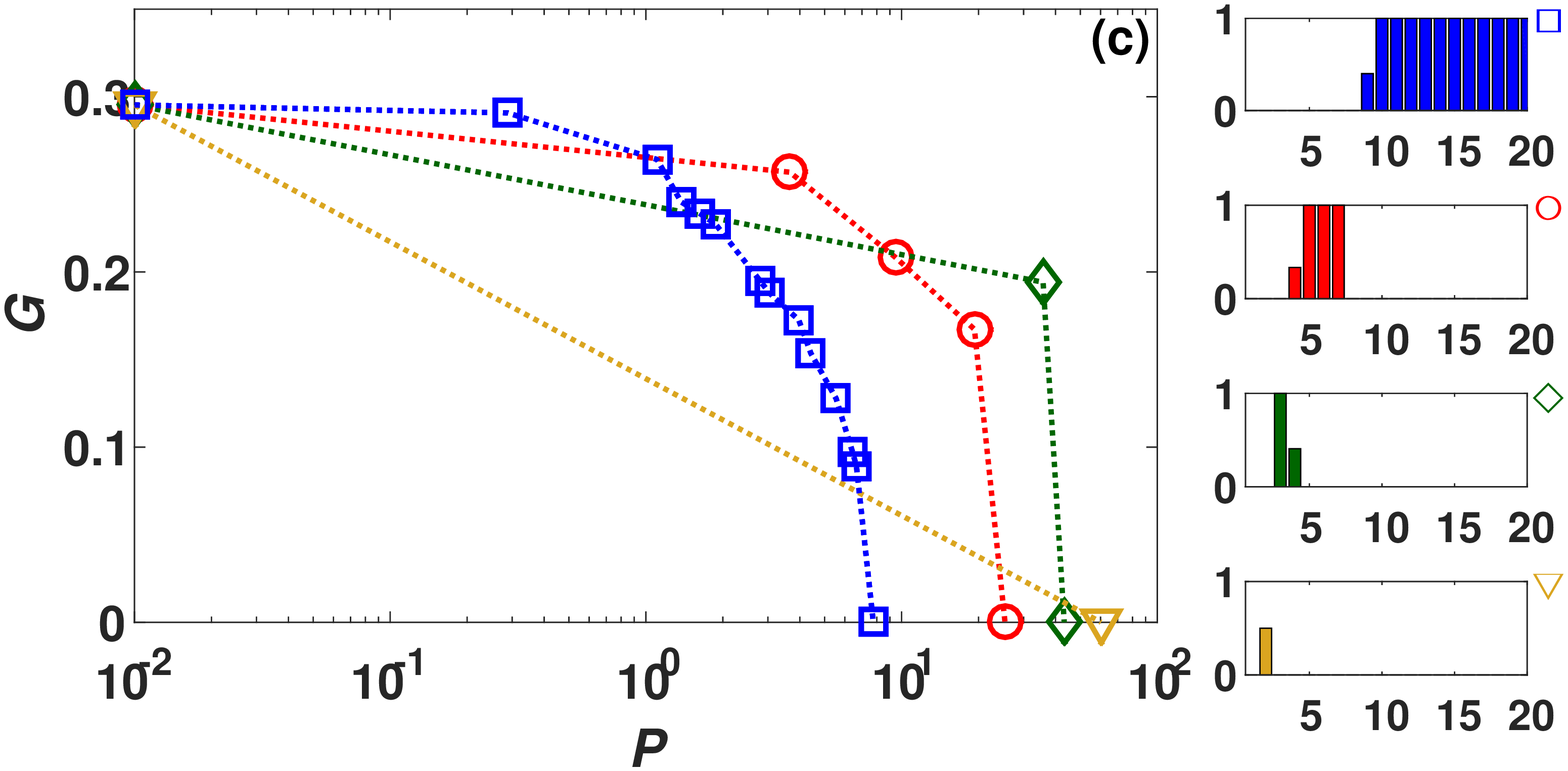}\hspace{-2.5em} & \includegraphics[width=10cm]{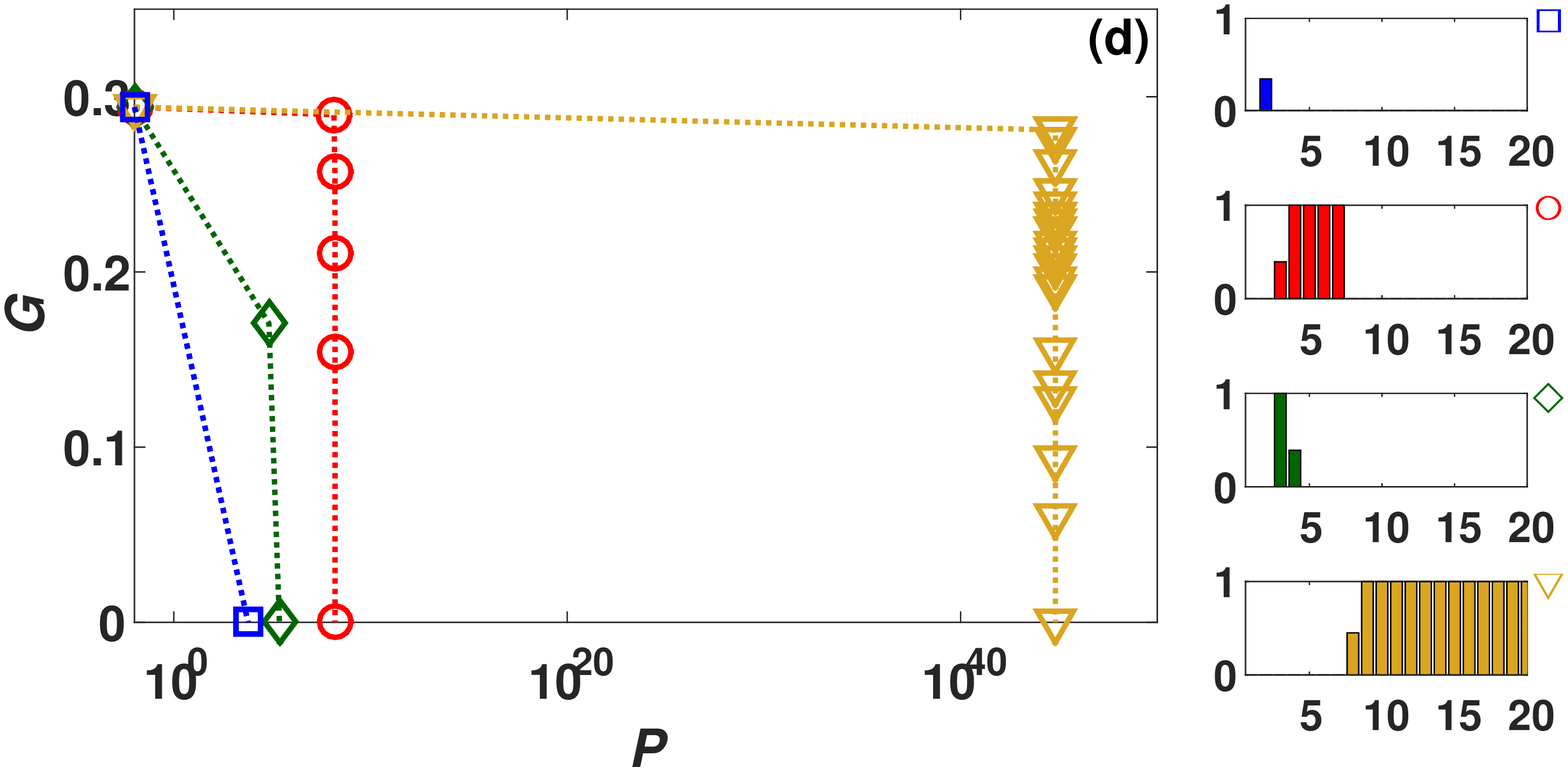} \\
	\end{tabular}
\end{center}
\caption{\textbf{Behavior of the size of the giant component in SF networks:} 
Open symbols represent simulations results of SF networks having $N=10^3$ nodes. The exponent $\gamma$ in the distribution $p(k)=Ck^{-\gamma}$ is $2.8$.
\textbf{(a,b)} Cost function $c(k)$ is power-law with:
\textbf{(a)} $\alpha=1.0$, \textbf{(b)} $\alpha=3.2$.
\textbf{(c,d)} Cost function $c(k)$ is exponential with:
\textbf{(c)} $\beta=0.01$, \textbf{(d)} $\beta=2.5$.
In each of the graphs the order of choosing the degrees to be removed, is identical to the graphs in Fig. \ref{fig:ER_G_vs_P}, except the curves that are signed by circles where only $1$ (and not $5$ as in Fig. \ref{fig:ER_G_vs_P}) percents of the nodes with the highest degrees are not removed. For the convenience, the bars in the right side of each figure were truncated arbitrarily above $k=20$, but in fact they include also the hubs with degrees up to about $k=40$.
Inset in \textbf{(b)}: Close up of the region where there are two curves that intersect the x-axis very closely. With accordance to our prediction, the curve that represents our method (rectangles) intersects the x-axis at the lowest point compared to the other curves.}
\label{fig:SF_G_vs_P}
\end{figure*}

An important property of Eq. (\ref{eq:z_k}) is the independence of $z(k)$ on $p(k)$. This means that our method is universal independent on the degree distribution of the special network we deal with. The only possible difference between various networks could be the stopping point of the chosen to be removed degrees. 
Accordingly, Fig. \ref{fig:SF_G_vs_P} illustrates results of simulations on SF networks. As can be seen, our optimal method establishes again a minimum cost of destroying the SF network, similar to the case of ER network.

\section{Efficient strengthening of network}
Strengthening a network in our method is the mirror image of the network's destruction model that presented above.   
We assume a network in which we are allowed to strengthen some of its nodes, such that at the beginning of an attack against the network all the nodes collapse except the nodes that were strengthened before. The strength of a node is measured by its survival time after an attack begun, that will be named the \emph{lifetime} of the node. We define a cost function $c(k)$ that is the cost of strengthening a node with degree $k$ by a lifetime of one unit of time. We classify the nodes by its degrees. Our goal is to find for every group of nodes with degree $k$, the fraction of nodes, that will be denoted by $q(k)$, to be strengthened by lifetime of one unit of time, such that the total cost of strengthening the entire network, which means to guarantee the existence of a giant component in the network, by lifetime of one unit of time is minimum.

We define a function $P$, that is the total cost of strengthening the entire network by lifetime of one unit, as follows
\begin{equation}
\label{eq:price_q}
P=\sum_{k=o}^\infty p(k)Nc(k)q(k)\enspace.
\end{equation}
The condition for percolation and the existence of a giant component is (see above Eq. (\ref{eq:threshold kappa}) and Eq. (\ref{eq:threshold expectation}))
\begin{equation}
\label{eq:perc_cond_strng}
\sum_{k=0}^\infty (k-1)\frac{kp(k)q(k)}{\lambda}=1\enspace,
\end{equation}
where for each $k$, $0\leq q(k)\leq1$. Note the similarity of Eq. (\ref{eq:price_q})-(\ref{eq:perc_cond_strng}), as well as the next equations to the analogous equations in the case of efficient destruction, but here we use $q(k)$ as opposed to $r(k)$ that we used in the previous case.

Very similar to the model of efficient destruction of network, by Eq. (\ref{eq:perc_cond_strng}) we define $a(k)$ as the contribution to the existence of a giant component of all the nodes with degree $k$ if all of them would be allocated by one unit of lifetime $\left(q\left(k\right)=1\right)$
\begin{equation}
a(k)\equiv (k-1)\frac{kp(k)}{\lambda}\enspace.
\end{equation}
By Eq. (\ref{eq:price_q}) we define $e(k)$ as the total cost of allocating all the nodes with degree $k$ by 1 unit of lifetime
\begin{equation}
e(k)\equiv p(k)c(k)N\enspace.
\end{equation}
$z(k)$ is again the ratio between $a(k)$ and $e(k)$ when neglecting the constants $\lambda$ and $N$, that is the ratio between the contribution of all the nodes with degree $k$ to the existence of the giant component and the cost of allocating all the nodes with degree $k$ by one unit of lifetime is,
\begin{equation}
\label{eq:z_k_strng}
\frac{a(k)}{e(k)}\propto\frac{(k-1)k}{c(k)}\equiv z(k)\enspace.
\end{equation}
Note the identity of $z(k)$ in Eq. (\ref{eq:z_k}) of the model of efficient destruction of network, and Eq. (\ref{eq:z_k_strng}) of the model of efficient strengthening a network. Like in the model of destruction of a network, we prefer to allocate lifetime to degrees with high value of $z(k)$.
We define an analogous method to that of efficient destroying of network, of how to strengthen a network with minimum cost, as follows: 

(i) For each degree $k$ calculate $z(k)$. 

(ii) Choose groups of degrees to be allocated by 1 unit of lifetime according to the value of $z(k)$ in descending order. 

The choice of degrees would be ended when a sufficient amount of degrees was chosen, such that if all of them would be allocated by one unit of lifetime the condition to percolation's threshold would be achieved, and a giant component with one unit of lifetime will appear in the network.

Although the identity of $z(k)$ between Eq. (\ref{eq:z_k}) of efficient destruction of network and Eq. (\ref{eq:z_k_strng}) of efficient strengthening of network, there is still a difference between the two cases regarding the critical threshold $p_c$ - the fraction of nodes that have to be functional to guarantee the existence of giant component in the network. Each of the two models is the mirror image of the other one. In the model of destruction of a network, we begin when all the nodes of the network are functional, then we destroy some nodes until we reach the percolation's threshold. In contrast in the model of strengthening a network, we begin when all the nodes are not functional, then we strengthen some nodes until we reach the percolation's threshold from the opposite direction. In the strengthening model we strengthen some degrees to construct from it the giant component, while in the destroying model we destroy exactly these degrees and construct the giant component from the other degrees. It is reasonable that in general when constructing a giant component from two different groups of degrees, $p_c$ is different.

\section{Summary}
In this work we developed a method for choosing the right group of nodes to be destroyed (immunized) or strengthened for minimizing the total price of destroying (immunizing) or strengthening a general random network. According to the value of a parameter $z(k)$ (Eq. (\ref{eq:z_k})), that we derived analytically, when calculated for each degree $k$, we define a list of priorities of degrees to be destroyed or strengthened, such that the cost for destroying or strengthening the entire network is minimum. Surprisingly, we find analytically that $z(k)$ is independent of the degree distribution $p(k)$, and therefore our method is general and useful for all kinds of random networks independent of the degree distribution of the network.

\vspace{15mm}

\renewcommand{\theequation}{A\arabic{equation}}
\setcounter{equation}{0}
\section{Appendix A:\\An analytic proof of the theory} 
We begin with the condition to percolation in a random network, that is
\begin{equation}
\sum_{k=0}^{\infty}(k-1)\frac{kp(k)q(k)}{\lambda}\leq1\enspace.
\label{eq:perc_cond}
\end{equation}
Substituting into it $q(k)=1-r(k)$ and $p(k)=\frac{N_k}{N}$, where $N_k$ is the expected number of nodes with degree $k$, we get
\begin{equation}
\sum_{k=0}^{\infty}(k-1)kN_kr(k)\geq\lambda N(\kappa-2)\enspace,
\label{eq:perc_cond_2}
\end{equation}
where $\kappa=\frac{\langle k^2\rangle}{\langle k\rangle}$. If we consider only one node with degree $k$ from all the $N_k$ nodes, we can see from Eq. (\ref{eq:perc_cond_2}) that its contribution to the percolation in the network is $r(k)(k-1)k$. 

Accordingly, we can replace the summation in Eq. (\ref{eq:perc_cond_2}) to be not over the degrees of the nodes, but over the nodes themselves, and get 
\begin{equation}
\sum_{v\in V}r(k)(k-1)k\geq\lambda N(\kappa-2)\enspace,
\label{eq:perc_cond_nodes}
\end{equation}
where $v$ is a specific node in the network and $V$ is the set of all the nodes in the network. The summation is calculated over all the nodes, and for each one of them with accordance to its degree $k$.

In our theory, $r(k)$ was determined according to the value of $z(k)$ (Eq. (\ref{eq:z_k})). We Choose degrees to be attacked (immunized) according to the value of $z(k)$ in descending order, and stopping the process when a sufficient combination of degrees are collected, such that if all these nodes are removed, the condition for percolation threshold would be achieved. Accordingly, all the degrees that were chosen are fully removed which for them $r(k)=1$, except the last degree that was chosen that usually is partially removed where $0<r(k)\leq1$. We can define a constant $M$ to be a threshold for $z(k)$, such that the set of nodes which for them $z(k)>M$ are fully removed, the set of nodes for which $z(k)=M$ are partially removed, and the set of nodes for which $z(k)<M$ are not chosen to be removed. Therefore, we can write the function $r(k)$ using the constant $M$ as follows
\begin{equation}
    r(k)= 
\begin{cases}
    1 & z(k)>M \\
	  \alpha & z(k)=M \enspace, \\
		0 & z(k)<M
\end{cases}
\label{eq:rk}
\end{equation}
where $0<\alpha\leq1$. To determine the specific value of $M$, we consider that in our theory we stop the process of removing nodes exactly when a percolation occurs and at the percolation threshold. Mathematically, that means that in the condition to percolation Eq. (\ref{eq:perc_cond_nodes}), among all the possibilities where the left hand side is greater or equals to the right-hand side, we choose a specific state where the two sides are equal. 
Therefeore, the value of $M$ has to be determined such that $r(k)$ fulfills the followings
\begin{equation}
\sum_{v\in V}r(k)(k-1)k=\lambda N(\kappa-2)\enspace.
\label{eq:perc_cond_nodes_eql}
\end{equation}

In the same manner that we replace the summation over degrees in Eq. (\ref{eq:perc_cond_2}) by a summation over nodes in Eq. (\ref{eq:perc_cond_nodes}), we can replace the summation over degrees in the function of the total cost to fragment the network Eq. (\ref{eq:price}) by a summation over nodes, and rewrite that function as follows
\begin{equation}
P=\sum_{v\in V}r(k)c(k)\enspace.
\label{eq:cost_nodes}
\end{equation}

Our theory argues that choosing nodes to be removed according to $r(k)$ Eq. (\ref{eq:rk}), where $M$ is determined according to the condition Eq. (\ref{eq:perc_cond_nodes_eql}), minimizes the cost function Eq. (\ref{eq:cost_nodes}).

We prove it as follows: assume an alternative way to $r(k)$ of choosing nodes to be removed, that will be named $r_A(k)$. The condition to percolation and fragmenting the network should be fulfilled also by removing nodes according to that function, such that the following condition is fulfilled
\begin{equation}
\sum_{v\in V}r_A(k)(k-1)k\geq\lambda N(\kappa-2)\enspace.
\label{eq:perc_cond_rA}
\end{equation}
 
We argue that the following inequality is true
\begin{equation}
\left[r_A(k)-r(k)\right]\left[k(k-1)-Mc(k)\right]\leq0\enspace.
\label{eq:inequality}
\end{equation}
We test its validity for three exhaustive options:\\
(i) $k(k-1)>Mc(k)$ which is equivalent to $z(k)>M$ -- the expression within the right parenthesis in Eq. (\ref{eq:inequality}) is positive. When $z(k)>M$, $r(k)=1$ and $r_A(k)\leq1$. Therefore, the expression within the left parenthesis is negative or equals to $0$. Thus, the multiplication of the two parenthesis in Eq. (\ref{eq:inequality}) is negative or equals to $0$ as we argued.\\
(ii) $k(k-1)=Mc(k)$ which is equivalent to $z(k)=M$ -- the expression within the right parenthesis in Eq. (\ref{eq:inequality}) equals to $0$. Therefore, Eq. (\ref{eq:inequality}) equals to $0$.\\
(iii) $k(k-1)<Mc(k)$ which is equivalent to $z(k)<M$ -- the expression within the right parenthesis in Eq. (\ref{eq:inequality}) is negative. When $z(k)<M$, $r(k)=0$ and $r_A(k)\geq0$. Therefore, the expression within the left parenthesis is positive or equals to $0$. Thus, the multiplication of the two parenthesis in Eq. (\ref{eq:inequality}) is negative or equals to $0$.

We rearrange Eq. (\ref{eq:inequality}) and also adding to it summation over all the nodes in the network, and get 
\begin{equation}
\sum_{v\in V}\left[r_A(k)-r(k)\right]k(k-1)\leq\sum_{v\in V} M\left[r_A(k)-r(k)\right]c(k)\enspace.
\label{eq:sum_inequality}
\end{equation}

By Eq. (\ref{eq:perc_cond_nodes_eql}) and Eq. (\ref{eq:perc_cond_rA}) we can see that the left-hand side of Eq. (\ref{eq:sum_inequality}) is greater or equals to $0$. Therefore, we get for the right-hand side of that equation 
\begin{equation}
\sum_{v\in V} M\left[r_A(k)-r(k)\right]c(k)\geq0\enspace,
\end{equation}
Thus we get 
\begin{equation}
\sum_{v\in V}r_A(k)c(k)\geq\sum_{v\in V}r(k)c(k)\enspace.
\end{equation}
By Eq. (\ref{eq:cost_nodes}) the former equation is equivalent to the following
\begin{equation}
P_A(k)\geq P(k)\enspace,
\label{eq:price_relation}
\end{equation}
Where $P_A(k)$ is the total price of destroying the network when choosing the nodes to be removed according to $r_A(k)$, and $P(k)$ is the total price of destroying the network when choosing the nodes according to $r(k)$ as suggested by our theory.
Therefore, Eq. (\ref{eq:price_relation}) shows that the total price of destroying the network according to $r(k)$ as suggested by our theory, is not greater than the total price of destroying the network when choosing the nodes according to any other alternative $r_A(k)$. Thus, we prove that choosing nodes to be removed according to our theory, minimizes the total price of destroying the network.

\renewcommand{\theequation}{B\arabic{equation}}
\setcounter{equation}{0}
\section{Appendix B: \\Analysis of $z(k)$ with power-law cost function} 
We begin with Eq. (\ref{eq:z_k}) and substituting into it $c(k)=k^\alpha$. We receive
\begin{equation}
z(k)=k^{1-\alpha}(k-1)\enspace.
\end{equation}
We differentiate it with respect to $k$, and receive
\begin{equation}
\frac{dz(k)}{dk}=k^{-\alpha}\left[\left(2-\alpha\right)k-\left(1-\alpha\right)\right]\enspace.
\label{eq: z_k_frstder}
\end{equation}
By zeroing this equation, we receive an extremum point - $k=\frac{1-\alpha}{2-\alpha}$, that would be denoted by $k_{ext}$.
We differentiate again and receive
\begin{equation}
\frac{d^2z(k)}{dk^2}=\left(\alpha-1\right)k^{-\alpha-1}\left[\left(\alpha-2\right)k-\alpha\right]\enspace.
\label{eq: z_k_secder}
\end{equation}
Substituting $k_{ext}$  into Eq. (\ref{eq: z_k_secder}), we receive
\begin{equation}
\frac{d^2z(k)}{dk^2}=\left(2-\alpha\right)\left(\frac{2-\alpha}{1-\alpha}\right)^\alpha\enspace.
\label{eq: z_k_secder_extrmpnt}
\end{equation}
For every $\alpha>2$ the second derivative in Eq. (\ref{eq: z_k_secder_extrmpnt}) is always negative, and thus $z(k)$ has a maximum point at $k_{ext}$. Note that $k_{ext}=2$ for $\alpha=3$, and as $\alpha$ increases $k_{ext}$ decreases until it equals to $1$ when $\alpha$ tends to infinity. Since our interest in $z(k)$ is only at $k\geq2$ (nodes with degrees $0$ or $1$ do not affect the destruction or the strengthening of the giant component), we conclude that as $\alpha\geq3$, $z(k)$ is a decreasing function for $k\geq2$.
When $\alpha<3$, $k_{ext}$ becomes greater than $2$, until $k_{ext}$ tends to infinity when $\alpha$ tends to $2^+$. Thus We conclude that for $2<\alpha<3$, $z(k)$ is an extremum maximum function. Despite that, note that when $k_{ext}$ is greater than the maximum degree of the network we analyze, $z(k)$ in fact becomes, for the sake of our problem, a monotonic increasing function. That especially as $\alpha$ tends to $2^+$ when $k_{ext}$ tends to infinity. 
 
In contrast, when $\alpha<1$ the second derivative in Eq. (\ref{eq: z_k_secder_extrmpnt}) is always positive, and thus $z(k)$ has a minimum point in $k_{ext}$. Recall that we are interested only in $\alpha>0$ (the cost function $k^\alpha$ is an increasing function with $k$), and that $k_{ext}$ tends to $0.5$ as $\alpha$ tends to $0^+$, and tends to $0$ as $\alpha$ tends to $1^-$, and also when $\alpha<1$ there is no discontinuity in $z(k)$ (Eq. (\ref{eq:z_k})) at $k=0$, we conclude that as $0<\alpha<1$, $z(k)$ is a monotonic increasing function for all $k\geq2$.

In the range $1\leq\alpha\leq2$, the analysis of an extremum points in $z(k)$ according to Eq. (\ref{eq: z_k_secder_extrmpnt}) is problematic, since we obtain complex numbers. However, we can do the analysis by using the fact that this range is the only one where both $(2-\alpha)$ and $-(1-\alpha)$ in the first derivative of $z(k)$ (Eq. (\ref{eq: z_k_frstder})) are positive. Thus in this range for all $k\geq0$, ans especially for every $k\geq2$, $z(k)$ always increases. We conclude that for $1\leq\alpha\leq2$, $z(k)$ is a monotonic increasing function for $k\geq2$.

In summary, the analysis of $z(k)$ with power-law cost function gives the following results at $k\geq2$:\\
1. When $0<\alpha\leq2$, $z(k)$ is a monotonic increasing function.\\
2. When $2<\alpha<3$, $z(k)$ is an extremum maximum function.\\
3. When $\alpha\geq3$, $z(k)$ is a monotonic decreasing function.\\

\renewcommand{\theequation}{C\arabic{equation}}
\setcounter{equation}{0}  
\section{Appendix C: \\Analysis of $z(k)$ with an exponential cost function} 
We begin with Eq. (\ref{eq:z_k}) and substituting into it $c(k)=e^{\beta k}$. We receive
\begin{equation}
z(k)=e^{-\beta k}k(k-1)\enspace.
\end{equation}
Differentiating it with respect to $k$, we receive
\begin{equation}
\frac{dz(k)}{dk}=e^{-\beta k}\left(-\beta k^2 +\beta k+2k-1\right)\enspace.
\label{eq: z_k_frstder2}
\end{equation}
By zeroing this equation, we receive two extremum points
\begin{equation}
k_1=\frac{\beta+2+\sqrt{\beta^2+4}}{2\beta}\qquad k_2=\frac{\beta+2-\sqrt{\beta^2+4}}{2\beta}\qquad\enspace.
\label{eq: kext2}
\end{equation}
Recall that $\beta>0$ (the cost function $e^{\beta k}$ is an increasing function with $k$). 
Analyzing $k_2$ we observe that it tends to $0.5$ as $\beta$ tends to $0^+$, and tends to $0$ as $\beta$ tends to $\infty$. Analyzing $k_1$ we observe that it tends to $\infty$ as $\beta$ tends to $0^+$, and tends to $1$ as $\beta$ tends to $\infty$. Note that $k_1>k_2$ for every $\beta$. Since we are interested in $z(k)$ only for $k\geq2$ and since the maximum of $k_2$ is $0.5$ less than $2$, then $z(k)$ for $k\geq2$ is affected only by $k_1$. Thus, we neglect $k_2$ and consider only $k_1$ that would be denoted by $k_{ext}$.
We differentiate $z(k)$ again and calculate the second derivative with $k=k_{ext}$. We receive 
\begin{equation}
\frac{d^2z(k)}{dk^2}=-\sqrt{\beta^2+4}\enspace,
\label{eq: z_k_secder_extrmpnt2}
\end{equation}
which is negative for every $\beta$, and thus $z(k)$ has a maximum point in $k_{ext}$ for every $\beta$.
From Eq. (\ref{eq: kext2}) it is easy to see that $k_{ext}=2$ when $\beta=1.5$. As $\beta$ increases above $1.5$, $k_{ext}$ decreases until $k_{ext}$ tends to $1$ as $\beta$ tends to infinity. On the other hand as $\beta$ decreases below $1.5$, $k_{ext}$ increases until $k_{ext}$ tends to infinity as $\beta$ tends to $0$. Recall that we only consider $z(k)$ at $k\geq2$, and thuswe conclude that as $\beta\geq1.5$, $z(k)$ is a monotonic decreasing function at $k\geq2$.  
On the other hand as $0<\beta<1.5$, $z(k)$ is an extremum maximum function at the range $k\geq2$. However, when $k_{ext}$ is greater than the maximum degree of the network we analyze, $z(k)$ in fact becomes a monotonic increasing function. That is especially valid as $\beta$ tends to $0^+$ where $k_{ext}$ tends to infinity.

In summary, the analysis of $z(k)$ with exponential cost function gives the following results for $k\geq2$:\\
1. When $0<\beta<1.5$, $z(k)$ is an extremum maximum function.\\
2. When $\beta\geq1.5$, $z(k)$ is a monotonic decreasing function.\\

\bibliography{minimum_cost_ver11}

\begin{thebibliography}{16}
\expandafter\ifx\csname natexlab\endcsname\relax\def\natexlab#1{#1}\fi
\expandafter\ifx\csname bibnamefont\endcsname\relax
  \def\bibnamefont#1{#1}\fi
\expandafter\ifx\csname bibfnamefont\endcsname\relax
  \def\bibfnamefont#1{#1}\fi
\expandafter\ifx\csname citenamefont\endcsname\relax
  \def\citenamefont#1{#1}\fi
\expandafter\ifx\csname url\endcsname\relax
  \def\url#1{\texttt{#1}}\fi
\expandafter\ifx\csname urlprefix\endcsname\relax\def\urlprefix{URL }\fi
\providecommand{\bibinfo}[2]{#2}
\providecommand{\eprint}[2][]{\url{#2}}

\bibitem[{\citenamefont{Erd\H{o}s and R{\'e}nyi}(1959)}]{erdds1959random}
\bibinfo{author}{\bibfnamefont{P.}~\bibnamefont{Erd\H{o}s}} \bibnamefont{and}
  \bibinfo{author}{\bibfnamefont{A.}~\bibnamefont{R{\'e}nyi}},
  \bibinfo{journal}{Publ. Math. Debrecen} \textbf{\bibinfo{volume}{6}},
  \bibinfo{pages}{290} (\bibinfo{year}{1959}).

\bibitem[{\citenamefont{Erd\H{o}s and R{\'e}nyi}(1960)}]{erd6s1960evolution}
\bibinfo{author}{\bibfnamefont{P.}~\bibnamefont{Erd\H{o}s}} \bibnamefont{and}
  \bibinfo{author}{\bibfnamefont{A.}~\bibnamefont{R{\'e}nyi}},
  \bibinfo{journal}{Publ. Math. Inst. Hungar. Acad. Sci}
  \textbf{\bibinfo{volume}{5}}, \bibinfo{pages}{17} (\bibinfo{year}{1960}).

\bibitem[{\citenamefont{Molloy and Reed}(1995)}]{molloy-randstruct-1995}
\bibinfo{author}{\bibfnamefont{M.}~\bibnamefont{Molloy}} \bibnamefont{and}
  \bibinfo{author}{\bibfnamefont{B.}~\bibnamefont{Reed}},
  \bibinfo{journal}{Random Structures \& Algorithms}
  \textbf{\bibinfo{volume}{6}}, \bibinfo{pages}{161} (\bibinfo{year}{1995}).

\bibitem[{\citenamefont{Albert and Barabasi}(2002)}]{albert-rev-mdn-phys-2002}
\bibinfo{author}{\bibfnamefont{R.}~\bibnamefont{Albert}} \bibnamefont{and}
  \bibinfo{author}{\bibfnamefont{A.-L.} \bibnamefont{Barabasi}},
  \bibinfo{journal}{Reviews of Modern Physics} \textbf{\bibinfo{volume}{74}},
  \bibinfo{pages}{47} (\bibinfo{year}{2002}).

\bibitem[{\citenamefont{Newman}(2010)}]{newman-networks-2010}
\bibinfo{author}{\bibfnamefont{M.}~\bibnamefont{Newman}},
  \emph{\bibinfo{title}{Networks: an introduction}} (\bibinfo{year}{2010}).

\bibitem[{\citenamefont{Albert et~al.}(2000)\citenamefont{Albert, Jeong, and
  Barab{\'a}si}}]{albert-nature-2000}
\bibinfo{author}{\bibfnamefont{R.}~\bibnamefont{Albert}},
  \bibinfo{author}{\bibfnamefont{H.}~\bibnamefont{Jeong}}, \bibnamefont{and}
  \bibinfo{author}{\bibfnamefont{A.-L.} \bibnamefont{Barab{\'a}si}},
  \bibinfo{journal}{Nature} \textbf{\bibinfo{volume}{406}},
  \bibinfo{pages}{378} (\bibinfo{year}{2000}).

\bibitem[{\citenamefont{Cohen et~al.}(2000)\citenamefont{Cohen, Erez,
  Ben-Avraham, and Havlin}}]{cohen-prl-2000}
\bibinfo{author}{\bibfnamefont{R.}~\bibnamefont{Cohen}},
  \bibinfo{author}{\bibfnamefont{K.}~\bibnamefont{Erez}},
  \bibinfo{author}{\bibfnamefont{D.}~\bibnamefont{Ben-Avraham}},
  \bibnamefont{and} \bibinfo{author}{\bibfnamefont{S.}~\bibnamefont{Havlin}},
  \bibinfo{journal}{Physical Review Letters} \textbf{\bibinfo{volume}{85}},
  \bibinfo{pages}{4626} (\bibinfo{year}{2000}).

\bibitem[{\citenamefont{Callaway et~al.}(2000)\citenamefont{Callaway, Newman,
  Strogatz, and Watts}}]{callaway-prl-2000}
\bibinfo{author}{\bibfnamefont{D.~S.} \bibnamefont{Callaway}},
  \bibinfo{author}{\bibfnamefont{M.~E.} \bibnamefont{Newman}},
  \bibinfo{author}{\bibfnamefont{S.~H.} \bibnamefont{Strogatz}},
  \bibnamefont{and} \bibinfo{author}{\bibfnamefont{D.~J.} \bibnamefont{Watts}},
  \bibinfo{journal}{Physical Review Letters} \textbf{\bibinfo{volume}{85}},
  \bibinfo{pages}{5468} (\bibinfo{year}{2000}).

\bibitem[{\citenamefont{Shuai et~al.}(2015)}]{shao-njp-2015}
\bibinfo{author}{\bibfnamefont{S.}~\bibnamefont{Shuai}} \bibnamefont{et~al.},
  \bibinfo{journal}{New Journal of Physics} \textbf{\bibinfo{volume}{17}},
  \bibinfo{pages}{023049} (\bibinfo{year}{2015}),
  \urlprefix\url{http://stacks.iop.org/1367-2630/17/i=2/a=023049}.

\bibitem[{\citenamefont{Cohen et~al.}(2001)\citenamefont{Cohen, Erez,
  Ben-Avraham, and Havlin}}]{cohen-prl-2001}
\bibinfo{author}{\bibfnamefont{R.}~\bibnamefont{Cohen}},
  \bibinfo{author}{\bibfnamefont{K.}~\bibnamefont{Erez}},
  \bibinfo{author}{\bibfnamefont{D.}~\bibnamefont{Ben-Avraham}},
  \bibnamefont{and} \bibinfo{author}{\bibfnamefont{S.}~\bibnamefont{Havlin}},
  \bibinfo{journal}{Physical Review Letters} \textbf{\bibinfo{volume}{86}},
  \bibinfo{pages}{3682} (\bibinfo{year}{2001}).

\bibitem[{\citenamefont{Gallos et~al.}(2005)}]{gallos-prl-2005}
\bibinfo{author}{\bibfnamefont{L.~K.} \bibnamefont{Gallos}}
  \bibnamefont{et~al.}, \bibinfo{journal}{Phys. Rev. Lett.}
  \textbf{\bibinfo{volume}{94}}, \bibinfo{pages}{188701}
  (\bibinfo{year}{2005}),
  \urlprefix\url{http://link.aps.org/doi/10.1103/PhysRevLett.94.188701}.

\bibitem[{\citenamefont{Schneider
  et~al.}(2011)}]{schneider-procacascience-2011}
\bibinfo{author}{\bibfnamefont{C.~M.} \bibnamefont{Schneider}}
  \bibnamefont{et~al.}, \bibinfo{journal}{Proceedings of the National Academy
  of Sciences} \textbf{\bibinfo{volume}{108}}, \bibinfo{pages}{3838}
  (\bibinfo{year}{2011}).

\bibitem[{\citenamefont{Achard et~al.}(2006)\citenamefont{Achard, Salvador,
  Whitcher, Suckling, and Bullmore}}]{achard-journal-neuroscince-2006}
\bibinfo{author}{\bibfnamefont{S.}~\bibnamefont{Achard}},
  \bibinfo{author}{\bibfnamefont{R.}~\bibnamefont{Salvador}},
  \bibinfo{author}{\bibfnamefont{B.}~\bibnamefont{Whitcher}},
  \bibinfo{author}{\bibfnamefont{J.}~\bibnamefont{Suckling}}, \bibnamefont{and}
  \bibinfo{author}{\bibfnamefont{E.}~\bibnamefont{Bullmore}},
  \bibinfo{journal}{The Journal of neuroscience} \textbf{\bibinfo{volume}{26}},
  \bibinfo{pages}{63} (\bibinfo{year}{2006}).

\bibitem[{\citenamefont{Li et~al.}(2013)}]{li-pre-2007}
\bibinfo{author}{\bibfnamefont{G.}~\bibnamefont{Li}} \bibnamefont{et~al.},
  \bibinfo{journal}{Phys. Rev. E} \textbf{\bibinfo{volume}{87}},
  \bibinfo{pages}{042810} (\bibinfo{year}{2013}),
  \urlprefix\url{http://link.aps.org/doi/10.1103/PhysRevE.87.042810}.

\bibitem[{\citenamefont{Morone and Makse}(2015)}]{morone-nature-2015}
\bibinfo{author}{\bibfnamefont{F.}~\bibnamefont{Morone}} \bibnamefont{and}
  \bibinfo{author}{\bibfnamefont{H.~A.} \bibnamefont{Makse}},
  \bibinfo{journal}{Nature} \textbf{\bibinfo{volume}{524}}, \bibinfo{pages}{65}
  (\bibinfo{year}{2015}).

\bibitem[{\citenamefont{Braunstein et~al.}(2016)\citenamefont{Braunstein,
  Dall’Asta, Semerjian, and Zdeborová}}]{Braunstein-pnas-2016}
\bibinfo{author}{\bibfnamefont{A.}~\bibnamefont{Braunstein}},
  \bibinfo{author}{\bibfnamefont{L.}~\bibnamefont{Dall’Asta}},
  \bibinfo{author}{\bibfnamefont{G.}~\bibnamefont{Semerjian}},
  \bibnamefont{and}
  \bibinfo{author}{\bibfnamefont{L.}~\bibnamefont{Zdeborová}},
  \bibinfo{journal}{Proceedings of the National Academy of Sciences}
  \textbf{\bibinfo{volume}{113}}, \bibinfo{pages}{12368}
  (\bibinfo{year}{2016}),
  \eprint{http://www.pnas.org/content/113/44/12368.full.pdf},
  \urlprefix\url{http://www.pnas.org/content/113/44/12368.abstract}.

\end{thebibliography}

\clearpage

\end{document}